\documentclass[
onecolumn, 11pt
]{IEEEtran}

\usepackage{amssymb}
\usepackage[cmex10]{amsmath}
\usepackage{float}
\usepackage[utf8]{inputenc}
\usepackage{cite}

\newtheorem{thm}{Theorem}
\newtheorem{lem}[thm]{Lemma}
\newtheorem{cor}[thm]{Corollary}
\newtheorem{defn}{Definition}

\newtheorem{rem}{Remark}
\newtheorem{claim}{Claim}

\newcommand{\supp}{\mathop{\mathrm{supp}}}
\newcommand{\tr}{{\mathop{\mathrm{tr}}}}
\newcommand{\Herm}{{\mathop{\mathrm{Herm}}}}
\newcommand{\app}[1]{~{\approx_{#1}}}
\newcommand{\id}{{\mathbb{I}}}
\newcommand{\Imaxb}{{^2I_{\max}\left(\rmA:\rmB\right)_\rho}}
\newcommand{\Imaxas}[1]{{^1I^{#1}_{\max}\left(\rmA:\rmB\right)_\rho}}
\newcommand{\Imaxbs}[1]{{^2I^{#1}_{\max}\left(\rmA:\rmB\right)_\rho}}
\newcommand{\Imaxbvars}[1]{{^2I^{#1}_{\max}\left(\rmB:\rmA\right)_\rho}}
\newcommand{\Imaxcs}[1]{{^3I^{#1}_{\max}\left(\rmA:\rmB\right)_\rho}}
\newcommand{\Hil}[1]{{\cal H}_{\rm #1}}
\newcommand{\Pos}{{\cal P}}
\newcommand{\Hmax}{{H_{\max}}}
\newcommand{\Hmin}{{H_{\min}}}
\newcommand{\Hmaxs}[1]{{H^{#1}_{\max}}}
\newcommand{\Hmins}[1]{{H^{#1}_{\min}}}
\newcommand{\BAB}[2]{{\cal B}^{#1}\left(#2\right)}
\newcommand{\eps}{\varepsilon}
\newcommand{\epsp}{{\varepsilon'}}
\newcommand{\rhop}{\rho'}
\newcommand{\rhoA}{\rho_\rmA}
\newcommand{\rhoB}{\rho_\rmB}
\newcommand{\rhoAB}{\rho_\rmAB}
\newcommand{\norm}[1]{\left\Vert{#1}\right\Vert}
\newcommand{\ket}[1]{{|{#1}\rangle}}
\newcommand{\ketbra}[1]{{|{#1}\rangle\langle{#1}|}}
\newcommand{\rendiv}[3]{{D_{#1}\left({#2}\Vert{#3}\right)}}
\newcommand{\smb}{{\sigma_\rmB\in S_=\left(\Hil{B}\right)}}
\newcommand{\sma}{{\sigma_\rmA\in S_=\left(\Hil{A}\right)}}
\newcommand{\defeq}{\mathrel{\mathop:}=}
\newcommand{\Seq}[1]{S_=\left({#1}\right)}
\newcommand{\Sleq}[1]{S_\leq\left({#1}\right)}
\newcommand{\rmA}{{\rm A}}
\newcommand{\rmB}{{\rm B}}
\newcommand{\rmC}{{\rm C}}
\newcommand{\rmAB}{{\rm AB}}
\newcommand{\da}{\Delta_\rmA^-}
\newcommand{\rmABC}{{\rm ABC}}
\newcommand{\Eps}{{\cal E}}

\title{Smooth Max-Information as One-Shot Generalization for Mutual Information}
\author{Nikola~Ciganovi\'c, Normand~J.~Beaudry, and~Renato~Renner
\thanks{This work was supported by the Swiss National Science Foundation (SNF) 
through the National Centre of Competence in Research “Quantum Science and Technology” 
(grant No. 200020-135048) and the European Research Council (grant 258932).
N.~J.~Beaudry, N.~Ciganovi\'c, and R.~Renner are with the Institute for Theoretical 
Physics, ETH Z\"urich, 8093 Z\"urich, Switzerland. (e-mail: nbeaudry@phys.ethz.ch; 
nikolac@student.ethz.ch; renner@phys.ethz.ch)}}

\begin{document}
\maketitle

\begin{abstract}
We study formal properties of smooth max-information, a generalization of von Neumann 
mutual information derived from the max-relative entropy. Recent work suggests that it 
is a useful quantity in one-shot channel coding, quantum rate distortion theory and 
the physics of quantum many-body systems.

Max-information can be defined in multiple ways. We demonstrate that different smoothed 
definitions are essentially equivalent (up to logarithmic terms in the smoothing 
parameters). These equivalence relations allow us to derive new chain rules for the 
max-information in terms of min- and max-entropies, thus extending the smooth entropy 
formalism to mutual information.
\end{abstract}


\section{Introduction}

\IEEEPARstart{M}{utual} information has been an important concept from the beginning of 
information theory. In classical information theory, the Shannon mutual 
information, 
\begin{equation}\label{ShannonInf}
  I(\rmA:\rmB)=H(\rmA)-H(\rmA|\rmB),
\end{equation} 
serves as a measure for the capacity of communication channels \cite{Sh48}. In quantum 
information theory, its analogue is given by the von Neumann mutual information which is 
defined in terms of von Neumann entropy in the same way as in (\ref{ShannonInf}). It generally 
presents a measure of correlation between the subsystems $\rmA$ and $\rmB$ of a composite 
quantum system. The operational drawback of these quantities from a practical point of view 
is that they only characterize processes under the assumption that they can be repeated an 
arbitrary number of times and that these repetitions are completely uncorrelated. In other 
words, the assumption states that the available resources are independent and identically 
distributed or {\it i.i.d.}. However, this assumption 
is not justified in more realistic settings. Channels for instance need
not be memoryless and the outputs for consecutive inputs may therefore be correlated. Also, assuming
an i.i.d. structure in cryptographic protocols may compromise their security since an adversary may perform an
attack that is not i.i.d.. A great amount of research has consequently been devoted to scenarios where the 
resources are not i.i.d., commonly called the {\it one-shot setting}. This scenario is not 
only closer to realistic communication settings, but can also be regarded as strictly more 
general. The i.i.d. case is a limiting case and can thus be reproduced from one-shot results. 
Hence, one-shot information theory also serves as a method for proving i.i.d. statements.

In order to characterize processes in the one-shot scenario,
the smooth min- and max-entropies $\Hmins{\eps}$ and $\Hmaxs{\eps}$ have been introduced 
\cite{RW04, Re05} and studied extensively both operationally and formally (see for example 
\cite{KRS09,TCR09,TCR10,Da09, To12, TSSR11}). They satisfy properties like 
data processing inequalities \cite{Re05, BR12} and a set of chain rules 
\cite{To12, Vi11}. Operationally, min- and max-entropies can be used to characterize 
various information theoretic tasks, including randomness extraction and state merging 
\cite{KRS09}. When the i.i.d.-limit is taken, {\it i.e.} if we evaluate them on average over 
$n$ for states of the 
form $\rho^{\otimes n}=\rho\otimes\rho\otimes\dots\otimes\rho$ with asymptotically large $n$, 
they indeed reproduce the von Neumann entropy \cite{TCR09,BR12} (this is called the quantum 
asymptotic equipartition property, or QAEP). Furthermore, smooth entropies have been shown 
to be asymptotically equivalent to an independent approach to non-asymptotic information 
theory \cite{DR08, Da09, TH12}, namely the information spectrum method as introduced by Han 
and Verd\' u in classical information theory \cite{HV93, VH94} and later generalized to the 
quantum setting by Nagaoka, Hayashi, Bowen, and Datta \cite{HN03, NH07, BD06}.
In light of the success of the smooth entropy formalism, the question arises of how 
it can be extended to mutual information in a meaningful way.

Recent research has produced a whole variety of expressions that appear to be useful one-shot 
generalizations for mutual information. Motivated by (\ref{ShannonInf}), generalized 
mutual information quantities can be defined as 
\begin{equation}\label{GenInf}
\begin{split}
  I^\eps_\text{gen}(\rmA:\rmB)&\defeq H_{\min}^\eps(\rmA)-H_{\min}^\eps(\rmA|\rmB)\\
  &\text{ or}~H_{\min}^\eps(\rmA)-H_{\max}^\eps(\rmA|\rmB)\\
  &\text{ or}~H_{\max}^\eps(\rmA)-H_{\max}^\eps(\rmA|\rmB)\\
  &\text{ or}~H_{\max}^\eps(\rmA)-H_{\min}^\eps(\rmA|\rmB).
\end{split}
\end{equation}
Several of these expressions have been found to have useful applications as bounds 
on one-shot capacities \cite{DH11, DHW13} or in the study of area laws in quantum 
statistical physics \cite{BH12}.

On the other hand, it is well known that the von Neumann entropy and mutual information can 
be defined as special cases of the quantum relative entropy 
$$\rendiv{}{\rho}{\sigma}=\tr\left(\rho(\log\rho-\log\sigma)\right),$$ 
where $\tr$ denotes the trace and $\log$ is the logarithm with base 2 throughout the paper. 
Therefore, it appears natural to define generalized information theoretic quantities in terms 
of generalized relative entropies. Min- and max-entropies for example are derived from 
the max- and min-relative entropy \cite{Da09}, 
\begin{equation*}
  \rendiv{\max}{\rho}{\sigma}=\min\{\lambda|2^{\lambda} \sigma\geq\rho\}
\end{equation*}
and
\begin{equation*} 
  \rendiv{\min}{\rho}{\sigma}=-\log\norm{\sqrt\rho\sqrt\sigma}_1^2,
\end{equation*}
respectively. In this paper, we focus on the mutual information quantity corresponding to 
the max-relative entropy, called the max-information. Recent work has established the max-
information as a relevant quantity in different information theoretic tasks. It has been 
identified by Berta {\it et al.} as a measure for the quantum communication cost of state 
splitting and state merging protocols \cite{BCR11, BRW13}. In addition, Datta {\it et al.} 
found the smooth max-information to characterize the minimal one-shot qubit compression 
size for a quantum rate distortion code \cite{DRRW13}. Apart from its information theoretic 
applications, it also appears to give a good characterization for the amount of correlation 
in spin systems \cite{Bla12}. However, there is again {\it a priori} no unique way in which 
such a quantity should be defined. Von Neumann mutual information can be defined in 
multiple ways in terms of the quantum relative entropy $\rendiv{}{\rho}{\sigma}$ \cite{BD10}, 
since
\begin{align}
  I(\rmA:\rmB)_\rho&=\rendiv{}{\rhoAB}{\rhoA\otimes\rhoB}\label{def1}\\
  &=\min_{\sigma_\rmB}\rendiv{}{\rhoAB}{\rhoA\otimes\sigma_\rmB}\label{asymm}\\
  &=\min_{\sigma_\rmA,\sigma_\rmB}\rendiv{}{\rhoAB}{\sigma_\rmA\otimes\sigma_\rmB}\label{def3},
\end{align}
where the minimizations run over all density operators $\sigma_\rmA$ and 
$\sigma_\rmB$ on $\Hil{A}$ and $\Hil{B}$, respectively. For other relative entropy measures 
these equalities do not hold in general. In fact, if we replace the quantum relative 
entropy with the max-relative entropy $\rendiv{\max}{\rho}{\sigma}$, the values of the three 
expressions above can lie arbitrarily far apart \cite{BCR11}: while the expressions of the form 
(\ref{asymm}) and (\ref{def3}) have a general upper bound given by $2\cdot\log\min\{|A|,|B|\}$, 
the expression of the form (\ref{def1}) is unbounded. Furthermore, the expression of the form 
(\ref{asymm}) is not symmetric in $\rmA$ and $\rmB$, unlike the von Neumann mutual 
information.

In order to consolidate and possibly unify these various approaches, it is of great 
interest to understand more about the relations among all these different quantities. In 
this paper, we 
show that smoothed versions of the max-information can be related to each other and 
regarded as approximately equivalent up to terms that depend only on the smoothing 
parameter and not on the specific quantum state or Hilbert space. These results can be 
employed to obtain chain rules in which we relate the max-information to differences of 
entropies as in (\ref{GenInf}). When evaluated for i.i.d.-states, these chain rules reproduce 
the well known relation (\ref{ShannonInf}) and thus imply the QAEP for the max-information. 
Since max-information and min-entropy are formally 
related via their definitions in terms of the max-relative entropy, we can adapt proof 
techniques from earlier work on min-entropy. 

The organization of the paper is as follows. In the next section we present the 
mathematical terminology and formal definitions necessary for the formulation of our 
results. Our results concerning the comparability of the different 
definitions and chain rules are summarized in sections \ref{ApprEquiv} and \ref{ChainRules}. 
Longer proofs, along with useful technical results, can be found in the appendices.

\section{Mathematical Preliminaries}\label{preliminaries}

\subsection{Basic Notations and Definitions}

In this paper we deal exclusively with finite dimensional Hilbert spaces $\Hil{A},\Hil{B}$ 
corresponding to physical systems $\rmA, \rmB$. To extend our results to infinite dimensional Hilbert spaces, the techniques of \cite{FAR11} could be used.
For tensor products of Hilbert spaces, we 
use the short notation $\Hil{AB}=\Hil{A}\otimes\Hil{B}$. Let $\Herm(\Hil{})$ be the space 
of Hermitian operators that act on $\Hil{}$ and $\Pos(\Hil{})\subseteq\Herm(\Hil{})$ the 
set of positive semi-definite operators on $\Hil{}$. For $A,B\in\Herm(\Hil{})$ we write 
$A\geq B$ iff $A-B\in\Pos(\Hil{})$. In this sense, we will sometimes write $A\geq 0$ in 
order to state that $A\in\Pos(\Hil{})$. The sets of normalized and subnormalized density operators on $\Hil{}$ are 
defined as 
$$\Seq{\Hil{}}\defeq\{\rho\in\Pos(\Hil{}):\tr\rho=1\}$$ 
and 
$$\Sleq{\Hil{}}\defeq\{\rho\in\Pos(\Hil{}):0<\tr\rho\leq 1\},$$ 
respectively. Operators are usually written with 
a subscript that specifies on which system they act, {\it e.g.} $\rhoAB\in\Herm(\Hil{AB})$. 
Given an operator $O_\rmAB$ on a composite Hilbert space $\Hil{AB}$, we obtain the reduced 
operator $O_\rmA$ on $\Hil{A}$ by taking the partial trace over the subsystem 
$\Hil{B}$: $O_\rmA=\tr_\rmB O_\rmAB$. The identity operator on $\Hil{A}$ is denoted by $\id_\rmA$.

Quantum operations are represented by completely positive and trace preserving (CPTP) maps, 
{\it i.e.} linear maps $\Eps:\Sleq{\Hil{}}\mapsto\Sleq{\Hil{}'}$ with the properties
$$\rho\geq 0 \Rightarrow \Eps(\rho)\geq 0,$$
and
$$\tr\rho=\tr\Eps(\rho),$$
for all $\rho\in\Sleq{\Hil{}}$. Note that the (partial) trace is a CPTP map.

Given any operator $O$, its operator norm $\norm{O}_{\infty}$ is given by its maximal 
singular value. Its trace norm is defined as $\norm{O}_1\defeq\tr\sqrt{O^\dagger O}$, where 
$O^\dagger$ is the adjoint of $O$. We will also require a notion of distance between 
density operators. For this purpose, we make use of the generalized fidelity, which is 
defined as 
\begin{equation*}
F(\rho,\sigma)\defeq\norm{\sqrt\rho\sqrt\sigma}_1+\sqrt{(1-\tr\rho)(1-\tr\sigma)},
\end{equation*}
for any $\rho,\sigma\in\Sleq{\Hil{}}$.
Note that when at least one of the states $\rho$ and $\sigma$ is normalized, 
\begin{equation*}
F(\rho,\sigma)=\norm{\sqrt{\rho}\sqrt{\sigma}}_1,
\end{equation*} 
which corresponds to the standard definition for fidelity. We use the fidelity to define a 
distance measure on $\Sleq{\Hil{}}$ as 
\begin{equation*}
P(\rho,\sigma)\defeq\sqrt{1-F^2(\rho,\sigma)}
\end{equation*} 
which is a metric (Lemma 5 in \cite{TCR10}). $P(\rho,\sigma)$ is called the {\it purified 
distance} between $\rho$ and $\sigma$. We say that two states $\rho$ and $\sigma$ are 
$\eps$-{\it close} and write $\rho\app{\eps}\sigma$ iff $P(\rho,\sigma)\leq\eps$.

Using the purified distance as a distance measure has many technical advantages. We 
summarize its essential properties, along with important properties of the fidelity in Appendix \ref{purdist}.

For any given $\rho\in\Sleq{\Hil{}}$, we can now define the ball of $\eps$-close states 
around $\rho$ as 
\begin{equation*}
\BAB{\eps}{\rho}\defeq\{\rhop\in\Sleq{\Hil{}}:P(\rho,\rhop)\leq\eps\},
\end{equation*} 
where $\eps$ is called the smoothing parameter and satisfies $0\leq\eps<\sqrt{\tr\rho}$, 
since we want to exclude the zero operator from the ball. In 
all of our statements, we make the implicit assumption that the involved smoothing 
parameters are small enough in this sense.

\subsection{Generalized Entropy Measures}

Let us now give the definitions for two types of generalized relative entropy, the max- and 
the min-relative entropy \cite{Da09}.
\begin{defn}
For $\rho,\sigma\in\Pos(\Hil{})$, the {\it max-relative entropy} is defined as
\begin{equation}
\rendiv{\max}{\rho}{\sigma}\defeq\min\{\lambda|2^{\lambda} \sigma\geq\rho\}.
\end{equation}
\end{defn}
Note that $\rendiv{\max}{\rho}{\sigma}$ to be well defined requires 
$\supp\rho\subseteq\supp\sigma$, where $\supp O$ denotes the support of the operator $O$, 
{\it i.e.} the space orthogonal to the kernel of $O$. If this is satisfied, 
there is an alternative way to express the max-relative entropy that we use 
frequently \cite{BCR11}: 
\begin{equation}\label{dmax:alt}
\rendiv{\max}{\rho}{\sigma}=\log\norm{\sigma^{-\frac{1}{2}}\rho\sigma^{-\frac{1}{2}}}_\infty.
\end{equation}
The inverses here are generalized inverses: given $\sigma\in\Pos(\Hil{})$, its generalized 
inverse $\sigma^{-1}$ is the unique minimum rank operator such that 
$\sigma^0\defeq\sigma\sigma^{-1}=\sigma^{-1}\sigma$ is the projector onto $\supp\sigma$.
\begin{defn}
For $\rho,\sigma\in\Pos(\Hil{})$, the {\it min-relative entropy of $\rho$ with respect 
to $\sigma$} is
\begin{equation} 
\rendiv{\min}{\rho}{\sigma}\defeq-\log\norm{\sqrt\rho\sqrt\sigma}_1^2.
\end{equation}
\end{defn}
Given any 
$\rhoAB\in\Sleq{\Hil{AB}}$, we can now define the (conditional) min- and max-entropies as 
$$\Hmin(\rmA|\rmB)_\rho\defeq -\min_{\smb}\rendiv{\max}{\rhoAB}{\id_\rmA\otimes\sigma_\rmB}$$ 
and 
$$\Hmax(\rmA|\rmB)_\rho\defeq -\min_{\smb}\rendiv{\min}{\rhoAB}{\id_\rmA\otimes\sigma_\rmB},$$ 
along with their smoothed versions: 
$$H_{\min}^\eps\left(\rmA|\rmB\right)_\rho\defeq\max_{\rhop\in\BAB{\eps}{\rho}}H_{\min}(\rmA|\rmB)_{\rhop},$$ 
and 
$$H_{\max}^\eps\left(\rmA|\rmB\right)_\rho\defeq\min_{\rhop\in\BAB{\eps}{\rho}}H_{\max}(\rmA|\rmB)_{\rhop}.$$ 
Min- and max-entropy are duals of each other in the sense that for pure $\rho_\rmABC$ \cite{TCR10} 
$$H^\eps_{\min}(\rmA|\rmB)_\rho=-H^\eps_{\max}(\rmA|\rmC)_\rho.$$ If the system $\rmB$ is trivial, 
we obtain the definitions for the non-conditional entropies: $$\Hmin(\rmA)_\rho=-\log\lambda_{\max}(\rhoA),$$ 
where $\lambda_{\max}(\rho)$ is the largest eigenvalue of $\rho$, while 
$$\Hmax(\rmA)_\rho=\log\norm{\sqrt\rhoA}_1^2$$ presents a measure for the fidelity between 
$\rhoA$ and the completely mixed state on $\Hil{\rmA}$.

\subsection{(Smooth) Max-Information}

As argued before, there is no unique way in which generalized mutual information measures 
should be obtained from the introduced relative entropies. Based on (\ref{def1})-(\ref{def3}), 
we define three different versions of max-information:
\begin{align*}
^1I_{\max}(\rmA:\rmB)_\rho&\defeq\rendiv{\max}{\rho_\rmAB}{\rho_\rmA\otimes\rho_\rmB},\\
^2I_{\max}(\rmA:\rmB)_\rho&\defeq\min_{\sigma_\rmB\in S_=\left(\Hil{B}\right)}\rendiv{\max}{\rho_\rmAB}{\rho_\rmA\otimes\sigma_\rmB},\\
^3I_{\max}(\rmA:\rmB)_\rho&\defeq\min_{\substack{\sigma_\rmA\in S_=\left(\Hil{A}\right),\\
\sigma_\rmB\in S_=\left(\Hil{B}\right)}}\rendiv{\max}{\rho_\rmAB}{\sigma_\rmA\otimes\sigma_\rmB}.
\end{align*}
For $\rho\in\Sleq{\Hil{AB}}$ and $\eps\geq 0$, we obtain {\it smooth max-information} from 
$^iI_{\max}\left(\rmA:\rmB\right)_\rho$ as 
\begin{equation*} 
^iI^\eps_{\max}\left(\rmA:\rmB\right)_\rho\defeq \min_{{\rho'}\in {\cal B}^\eps\left(\rho\right)}{^iI_{\max}\left(\rmA:\rmB\right)_{\rho'}}.
\end{equation*} 
It should be pointed out that earlier literature making use of smooth max-information 
usually refers to $\Imaxbs{\eps}$. In particular, a chain rule, a data processing 
inequality and the QAEP have been proven for $^2I_{\max}$ in \cite{BCR11}. The proof 
of the data processing inequality can straightforwardly be extended to all smooth definitions.

\begin{lem}
Let $\rhoAB\in\Sleq{\Hil{AB}}$, $\eps\geq 0$ and let $\Eps$ be a CPTP map of the form $\Eps=\Eps_\rmA\otimes\Eps_\rmB$. Then
\begin{equation}
^iI_{\max}^{\eps}(\rmA:\rmB)_{\Eps(\rho)}\leq {^iI_{\max}^{\eps}}(\rmA:\rmB)_\rho,
\end{equation}
for any $i\in\{1,2,3\}$.
\end{lem}

\begin{IEEEproof}
We provide the proof for $i=2$, the other cases being similar. Let 
$\rho'_\rmAB\in\BAB{\eps}{\rhoAB}$ be a state that optimizes $\Imaxbs{\eps}$, 
{\it i.e.} $\Imaxbs{\eps}={^2I_{\max}}(\rmA:\rmB)_{\rhop}$. Then there exists $\smb$ 
such that 
\begin{align*}
\Imaxbs{\eps}&=\rendiv{\max}{\rho'_\rmAB}{\rho'_\rmA\otimes\sigma_\rmB}\\
&\geq\rendiv{\max}{\Eps(\rho'_\rmAB)}{\Eps_\rmA(\rho'_\rmA)\otimes\Eps_\rmB(\sigma_\rmB)}\\
&\geq\min_{\omega_\rmB\in\Seq{\Hil{B}}}\rendiv{\max}{\Eps(\rho'_\rmAB)}{\Eps_\rmA(\rho'_\rmA)\otimes\omega_\rmB}\\
&\geq\min_{\substack{\bar\rho\in\BAB{\eps}{\Eps(\rho)},\\\omega_\rmB\in\Seq{\Hil{B}}}}\rendiv{\max}{\bar\rho_\rmAB}{\bar\rho_\rmA\otimes\omega_\rmB},
\end{align*}
where the first inequality follows from the data processing inequality for the 
max-relative entropy (cf. Lemma \ref{datproc}) and the last inequality is a consequence 
of the monotonicity of the purified distance under trace non-increasing CPMs (cf. Lemma \ref{prel12}).
\end{IEEEproof}


\section{Approximate Equivalence Relations for $^iI_{\max}^\eps$}\label{ApprEquiv}

Let us now turn to our main problem of relating alternative expressions for smooth 
max-information to each other. Our key results are given by the following two theorems. 
For convenience of notation, we introduce the two functions
$$f(\eps,\epsp)\defeq\log\left({\frac{1}{1-\sqrt{1-\eps^2}}}+\frac{1}{1-\epsp}\right)$$
and
$$g(\eps)\defeq \log\left({\frac{2(1-\eps)+3}{(1-\eps)(1-\sqrt{1-\eps^2})}}\right).$$
Note that both functions grow logarithmically in $\frac{1}{\eps}$ as $\eps\rightarrow 0$.

\begin{thm}\label{equiv:1}
Let $\rhoAB\in S_=\left(\Hil{AB}\right)$ and $\eps>0$, $\epsp\geq 0$. Then 
\begin{equation}
\begin{split}
\Imaxcs{\eps+\epsp}&\leq\Imaxbs{\eps+\epsp}\\
&\leq\Imaxcs{\epsp}+f(\eps,\epsp).
\end{split}
\end{equation}
\end{thm}

\begin{thm}\label{equiv:2}
Let $\rhoAB\in S_=\left(\Hil{AB}\right)$ and $\eps>0$, $\epsp\geq 0$. Then, 
\begin{equation}
\begin{split}
\Imaxbs{\eps+2\sqrt{\eps}+\epsp}&\leq\Imaxas{\eps+2\sqrt{\eps}+\epsp}\\
&\leq\Imaxbs{\epsp}+g(\eps).
\end{split}
\end{equation}
\end{thm}

We provide the proofs of these theorems in the appendix and turn immediately to the 
corollaries. First we complete our set of approximate equivalence relations. In order to 
compare $^1I_{\max}^\eps$ and $^3I_{\max}^\eps$, we only need to combine Theorems 
\ref{equiv:1} and \ref{equiv:2}. 

\begin{cor}\label{equiv:3}
Let $\rhoAB\in S_=\left(\Hil{AB}\right)$ and $\eps,\epsp>0$, $\eps''\geq 0$. Then
\begin{equation}
\begin{split}
\Imaxcs{\eps+2\sqrt{\eps}+\epsp+\eps''}\leq{}&\Imaxas{\eps+2\sqrt{\eps}+\epsp+\eps''}\\
\leq{}&\Imaxcs{\eps''}\\
&+f(\eps',\eps'')+g(\eps).
\end{split}
\end{equation}
\end{cor}

We thus conclude that all three definitions for $I_{\max}^\eps$ are pairwise approximately 
equivalent, meaning that since the differences between them are independent of the given 
state or Hilbert space, they must carry the same qualitative content. 

These relations further imply an estimate on the approximate symmetry of $^2I^\eps_{\max}$.

\begin{cor}\label{equiv:4}
Let $\rhoAB\in S_=\left(\Hil{AB}\right)$ and $\eps>0$, $\epsp\geq 0$. Then
\begin{equation} 
\begin{split}
\Imaxbs{2\eps+\epsp}&\leq\Imaxbvars{\eps+\epsp}+f(\eps,\eps+\epsp)\\
&\leq\Imaxbs{\epsp}+f(\eps,\eps+\epsp)+f(\eps,\epsp).
\end{split}
\end{equation}
\end{cor}

\begin{IEEEproof}
Note that $\Imaxbvars{\eps}\geq\Imaxcs{\eps}$, which follows directly from the definitions. 
Then, using Theorem~\ref{equiv:1} and the apparent symmetry of $^3I_{\max}^\eps$, we find 
that
\begin{equation*}
\Imaxbvars{\eps+\epsp}\leq\Imaxbs{\epsp}+f(\eps,\epsp),
\end{equation*}
as well as
\begin{equation*}
\Imaxbs{\eps+\epsp}\leq\Imaxbvars{\epsp}+f(\eps,\epsp),
\end{equation*}
and the claim follows.
\end{IEEEproof}

\section{Chain Rules for $^iI^\eps_{\max}$}\label{ChainRules}
In this section we prove chain rules for smooth max-information of the form
\begin{align*} 
H^\eps_{\min}(\rmA)_\rho-H^\eps_{\min}(\rmA|\rmB)_\rho &\lesssim I_{\max}^{\eps}(\rmA:\rmB)_\rho\\
&\lesssim H_{\max}^\eps(\rmA)_\rho-H_{\min}^\eps(\rmA|\rmB)_\rho.
\end{align*}
An upper bound chain rule for $^2I_{\max}^\eps$ is already known from Lemma B.15 in 
\cite{BCR11}: for $\rho\in\Seq{\Hil{}}$ and $\eps>0$,
\begin{equation}\label{chainup}
\begin{split}
\Imaxbs{\eps}\leq H_{\max}^{\eps^2/48}(\rmA)_\rho-H_{\min}^{\eps^2/48}(\rmA|\rmB)_\rho-l(\eps),
\end{split}
\end{equation}
where $l(\eps)\defeq 2\cdot\log\frac{\eps^2}{24}$. Let us first derive a lower bound chain 
rule for $^2I_{\max}^\eps$. Having both bounds for one of the definitions will allow us to 
write down chain rules for all $^iI_{\max}^\eps$ by exploiting the approximate equivalence 
relations from the previous section.

\begin{lem}\label{chainlow}
Let $\rho\in\Seq{\Hil{AB}}$ and $\eps\geq 0$. Then 
\begin{equation}
\Imaxbs{\eps}\geq H_{\min}^{\eps}(\rmA)_\rho-H_{\min}^{4\sqrt{2\eps}}(\rmA|\rmB)_\rho.
\end{equation}
\end{lem}

\begin{IEEEproof}
The proof is similar to the one of (\ref{chainup}) in \cite{BCR11}. We rearrange Lemma B.13 
from \cite{BCR11} as 
\begin{equation}
H_{\min}(\rmA|\rmB)_\rho\geq H_{\min}(\rmA)_\rho-\Imaxb.
\end{equation} 
Thus, 
\begin{align*}
&H_{\min}^\eps(\rmA|\rmB)_\rho\\
\geq{}&\max_{\rhop\in\BAB{\eps}{\rho}}\left[H_{\min}(\rmA)_{\rhop}-{^2I}_{\max}(\rmA:\rmB)_{\rhop}\right]\\
\geq{}&\max_{\omega\in\BAB{\eps^2/32}{\rho}}\max_{\Pi_\rmA}\left[H_{\min}(\rmA)_{\Pi\omega\Pi}-{^2I}_{\max}(\rmA:\rmB)_{\Pi\omega\Pi}\right],
\end{align*}
where the maximization runs over all 
$0\leq\Pi_\rmA\leq\id_\rmA$ with $\Pi_\rmA\omega\Pi_\rmA\app{\eps/2}\omega$. 
Next, choose $\omega'\in\BAB{\eps^2/32}{\rho}$ such that 
$^2I_{\max}(\rmA:\rmB)_{\omega'}=\Imaxbs{\eps^2/32}$. This gives us 
\begin{align*} 
H_{\min}^\eps(\rmA|\rmB)_\rho&\geq\max_{\Pi_\rmA}\left[H_{\min}(\rmA)_{\Pi\omega'\Pi}-{^2I}_{\max}(\rmA:\rmB)_{\Pi\omega'\Pi}\right]\\
&\geq\max_{\Pi_\rmA}H_{\min}(\rmA)_{\Pi\omega'\Pi}-{^2I}_{\max}(\rmA:\rmB)_{\omega'},
\end{align*}
with the maximization running over all $0\leq\Pi_\rmA\leq\id_\rmA$ with 
$\Pi_\rmA\omega'\Pi_\rmA\app{\eps/2}\omega'$. The second inequality is a consequence of 
Remark~\ref{maxmon} (cf. Appendix \ref{techlemmas}). According to Lemma~\ref{tech2}, we can 
choose a $\Pi_\rmA$ such that 
$H_{\min}^{\eps^2/16}(\rmA)_{\omega'}\leq H_{\min}(\rmA)_{\Pi\omega'\Pi}$. Doing so yields
\begin{align*} 
H_{\min}^\eps(\rmA|\rmB)_\rho&\geq H_{\min}^{\eps^2/16}(\rmA)_{\omega'}-\Imaxbs{\eps^2/32}\\
&\geq H_{\min}^{\eps^2/32}(\rmA)_\rho-\Imaxbs{\eps^2/32}.
\end{align*}
Relabelling $\eps^2/32\rightarrow\eps$ concludes the proof.
\end{IEEEproof}

We can now obtain chain rules for alternative definitions of $I_{\max}^\eps$ as well.

\begin{cor}
Let $\rho\in\Seq{\Hil{AB}}$ and $\eps,\epsp>0$. Then 
\begin{equation}
\begin{split}
\Imaxas{\eps+2\sqrt{\eps}+\epsp}\leq {}&H_{\max}^{\epsp^2/48}(\rmA)_\rho-H_{\min}^{\epsp^2/48}(\rmA|\rmB)_\rho\\
&+g(\eps)-l(\epsp).
\end{split}
\end{equation}
\end{cor}
\begin{IEEEproof}
With (\ref{chainup}) and using Theorem~\ref{equiv:2} to estimate 
$\Imaxas{\eps+2\sqrt\eps+\epsp}$ in terms of $\Imaxbs{\epsp}$, the claim follows immediately.
\end{IEEEproof}

Similarly, we obtain a lower bound chain rule for $\Imaxcs{\eps}$:

\begin{cor}
For $\rho\in\Seq{\Hil{AB}}$ and $\eps>0$, $\epsp\geq 0$,
\begin{equation}
\begin{split}
\Imaxcs{\epsp}\geq {}&H_{\min}^{\eps+\epsp}(\rmA)_\rho-H_{\min}^{4\sqrt{2\eps+2\epsp}}(\rmA|\rmB)_\rho\\&-f(\eps,\epsp).
\end{split}
\end{equation}
\end{cor}

\begin{IEEEproof}
The claim is a direct consequence of Theorem \ref{equiv:1} and Lemma \ref{chainlow}.
\end{IEEEproof}

\section{Conclusion}

We have investigated properties of smooth max-information defined as a special case of the 
max-relative entropy. In earlier work, it has been shown to be an operational quantity in 
one-shot state splitting and state merging \cite{BCR11, BRW13}. It is also found to be a 
useful quantity in quantum rate distortion theory \cite{DRRW13} and the statistical physics 
of many body systems \cite{Bla12}. We have shown that it exhibits some properties that we 
would expect from previous results on smooth 
entropies. Alternative definitions of max-information turn out to be essentially equivalent. 
Furthermore, they satisfy upper and lower bound chain rules in terms of min- and max-entropies.
Chain rules are generally an important technical tool in information theory. In this case, 
they also relate max-information to alternative definitions for one-shot mutual 
information, made up from differences of entropies as used in \cite{DH11, DHW13, BH12}. 

The primary goal of further research on these quantities is to gain a better 
understanding of their operational relevance. We hope that the formal tools provided in 
this paper will be useful for this purpose.

\section*{Acknowledgments} NC would like to thank Joseph M. Renes and Frédéric Dupuis 
for helpful discussions.


\appendices
\renewcommand{\theequation}{\thesection.\arabic{equation}}
\section{Properties of the Fidelity and Purified Distance}\label{purdist}

Here we summarize the essential properties of the purified distance. For a more 
extensive discussion, we refer the reader to \cite{To12}. 
The main reasons the purified distance is preferred over the trace distance are 
the following two lemmas, which state that for given $\rho, \sigma$, we can always 
find purifications or extensions $\bar{\rho},\bar{\sigma}$ such that 
$P(\rho,\sigma)=P(\bar\rho,\bar\sigma)$. This is due to Uhlmann's theorem \cite{Uhl85}: 
for any states $\rhoA,\sigma_\rmA\in\Seq{\Hil{A}}$ 
\begin{equation}
\norm{\sqrt{\rhoA}\sqrt{\sigma_\rmA}}_1=\max_{\rhoAB,\sigma_\rmAB}\norm{\sqrt{\rhoAB}\sqrt{\sigma_\rmAB}}_1,
\end{equation} 
where  the maximization runs over all purifications $\rhoAB$ and 
$\sigma_\rmAB$ of $\rhoA$ and $\sigma_\rmA$. 
\begin{lem}[Lemma 8 in \cite{TCR10}]\label{prel7}
Let $\rho,\sigma\in\Sleq{\Hil{}}$, $\Hil{}'\cong\Hil{}$ and 
$\varphi\in\Hil{}\otimes\Hil{}'$ be a purification of $\rho$. Then there exists 
a purification $\vartheta\in\Hil{}\otimes\Hil{}'$ of $\sigma$ with 
$P(\rho,\sigma)=P(\varphi,\vartheta)$.
\end{lem}

\begin{lem}[Corollary 9 in \cite{TCR10}]\label{prel8}
Let $\rho,\sigma\in\Sleq{\Hil{}}$ and $\bar\rho\in\Sleq{\Hil{}\otimes\Hil{}'}$ be 
an extension of $\rho$. Then there exists an extension 
$\bar\sigma\in\Sleq{\Hil{}\otimes\Hil{}'}$ of $\sigma$ with 
$P(\rho,\sigma)=P(\bar\rho,\bar\sigma)$.
\end{lem}

Still, the purified distance is equivalent to the generalized trace distance given by
$$D(\rho,\sigma)=\frac{1}{2}\norm{\rho-\sigma}_1+\frac{1}{2}|\tr\rho-\tr\sigma|.$$
Therefore it retains an operational interpretation as a measure for the maximum guessing 
probability \cite{NC00}: the maximal probability $p_{\mathrm{dist}}(\rho,\sigma)$ for 
correctly distinguishing between two quantum states $\rho,\sigma$ satisfies 
$$p_{\mathrm{dist}}(\rho,\sigma)\leq\frac{1}{2}(1+D(\rho,\sigma)).$$

\begin{lem}[Lemma 7 in \cite{TCR10}]
Let $\rho,\sigma\in\Sleq{\Hil{}}$. Then
\begin{equation}\label{trace::pure}
D(\rho,\sigma)\leq P(\rho,\sigma)\leq \sqrt{2D(\rho,\sigma)}.
\end{equation}
\end{lem}

Another useful property of the purified distance is that it cannot increase under trace 
non-increasing CPMs.

\begin{lem}[Lemma 7 in \cite{TCR10}]\label{prel12}
Let $\rho,\sigma\in\Sleq{\Hil{}}$ and let ${\cal E}$ be a trace non-increasing CPM. 
Then $P(\rho,\sigma)\geq P({\cal E}(\rho),{\cal E}(\sigma)).$
\end{lem}

We make use of the following properties of the standard fidelity.

\begin{lem}[\cite{To12}]\label{fid}
Let $\rho,\sigma\in\Pos(\Hil{})$. 
\begin{itemize}
\item For any $\omega\geq\rho$, 
\begin{equation}
\norm{\sqrt\omega\sqrt\sigma}_1\geq\norm{\sqrt\rho\sqrt\sigma}_1.
\end{equation} 
\item For any projector $\Pi\in\Pos(\Hil{})$, 
\begin{equation}\label{fid3}
\begin{split}
\norm{\sqrt{\Pi\rho\Pi}\sqrt{\sigma}}_1&=\norm{\sqrt{\rho}\sqrt{\Pi\sigma\Pi}}_1\\
&=\norm{\sqrt
{\Pi\rho\Pi}\sqrt{\Pi\sigma\Pi}}_1.
\end{split}
\end{equation}
\end{itemize}
\end{lem}

We conclude this section by stating a few useful technical facts.

\begin{lem}[Lemma 17 in \cite{TSSR11}]\label{prel9}
Let $\rho\in\Sleq{\Hil{}}$ and $\Pi$ a projector on $\Hil{}$, then 

\begin{equation}
P(\rho,\Pi\rho\Pi)\leq\sqrt{2\cdot\tr(\Pi^\bot\rho)-\left(\tr(\Pi^\bot\rho)\right)^2},
\end{equation} 
where $\Pi^\bot=\id-\Pi$.
\end{lem}

\begin{lem}[Lemma A.7 in \cite{BCCRR10}]\label{prel10}
Let $\rho\in\Sleq{\Hil{}}$ and $\Pi\in\Pos({\Hil{}})$ such that $\Pi\leq\id$. Then 
\begin{equation}
P(\rho,\Pi\rho\Pi)\leq\frac{1}{\sqrt{\tr\rho}}\sqrt{\left(\tr(\rho)\right)^2-\left(\tr(\Pi^2\rho)\right)^2}.
\end{equation}
\end{lem}

\begin{cor}\label{prel11}
Let $\rho\in\Sleq{\Hil{}}$ and $0<k\leq 1$. Then 
\begin{equation}
P(\rho,k\cdot\rho)\leq\sqrt{1-k^2}.
\end{equation}
\end{cor}
\begin{IEEEproof}
Apply Lemma~\ref{prel10} to $\Pi=\sqrt k\cdot\id$ and use $\sqrt{\tr\rho}\leq 1$.
\end{IEEEproof}

\section{Technical Lemmas}\label{techlemmas}

The following lemma introduces a notion of duality between projectors on subsystems of a 
multi-partite quantum system with respect to a given pure state. It is essential in the 
proofs of Theorems \ref{equiv:1} and \ref{equiv:2}.
\begin{lem}[Corollary 16 in \cite{TSSR11}]\label{prel5}
Let $\rho_\rmAB=\ketbra{\varphi}_\rmAB\in\Pos(\Hil{AB})$ be pure, 
$\rhoA=\tr_\rmB\rho_\rmAB$, $\rhoB=\tr_\rmA\rho_\rmAB$ and let $\Pi_\rmA\in\Pos(\Hil{A})$ 
be a projector in $\supp\rhoA$. Then, there exists a dual projector $\Pi_\rmB$ on $\Hil{B}$ 
such that 
\begin{equation} 
(\Pi_\rmA\otimes\rhoB^{-1/2})\ket{\varphi}_\rmAB=(\rhoA^{-1/2}\otimes\Pi_\rmB)\ket{\varphi}_\rmAB.
\end{equation}
\end{lem}

The proof of Theorem \ref{equiv:2} further requires the following inequality for the 
operator norm.
\begin{lem}\label{simple}
Let $A, B, C\in\Pos(\Hil{})$ be such that $\supp A\subseteq\supp B$ and $B\leq C$. Then  
\begin{equation}
\norm{C^{-1/2}AC^{-1/2}}_\infty\leq\norm{B^{-1/2}AB^{-1/2}}_\infty.
\end{equation}
\end{lem}

\begin{IEEEproof}
We know from (\ref{dmax:alt}) that $\lambda=\norm{B^{-1/2}AB^{-1/2}}_\infty$ is the 
smallest number such that $A\leq\lambda B$. Then $B\leq C$ implies $A\leq\lambda C$ 
and the claim follows.
\end{IEEEproof}

In proving the chain rules for $I_{\max}$, we have used the following facts on different 
entropic quantities.

\begin{lem}[Lemma 5 in \cite{On12}]\label{tech2}
For any $\rho\in\Sleq{\Hil{A}}$ and $\eps\geq 0$, there exists an operator 
$0\leq\Pi\leq\id_\rmA$ such that $\rho\app{\eps/2}\Pi\rho\Pi$ and 
\begin{equation} 
H_{\min}^{\eps^2/16}(\rmA)_\rho\leq H_{\min}(\rmA)_{\Pi\rho\Pi}.
\end{equation}
\end{lem}

\begin{lem}[Lemma 7 in \cite{Da09}]\label{datproc}
Let $\rho,\sigma\in\Pos(\Hil{})$ and $\cal E$ be a CPTP map on $\Hil{}$. Then 
\begin{equation} 
\rendiv{\max}{\rho}{\sigma}\geq\rendiv{\max}{{\cal E}(\rho)}{{\cal E}(\sigma)}.
\end{equation}
\end{lem}

\begin{rem}\label{maxmon}
This actually holds more generally even if the CPM $\cal E$ is not trace preserving. 
In particular, for any $\Pi\in\Pos(\Hil{})$, 
\begin{equation} 
\rendiv{\max}{\rho}{\sigma}\geq\rendiv{\max}{\Pi\rho\Pi}{\Pi\sigma\Pi}.
\end{equation}
\end{rem}

\section{Proofs of Theorems \ref{equiv:1} and \ref{equiv:2}}
\subsection{Auxiliary Lemmas}
Before turning to the main proofs, we want to make a few observations on the normalization 
of optimal operators for $^iI_{\max}^\eps$. Lemma~\ref{30} will prove especially useful in 
the proof of Thereom~\ref{equiv:2}.

The proofs of these lemmas rely on the following fact.

\begin{lem}\label{10}
Let $\rhoAB\in\Seq{\Hil{AB}}$, $\eps\geq 0$ and $\rhop\in\BAB{\eps}{\rhoAB}$. Then 
$\frac{\rhop}{\tr\rhop}\in\BAB{\eps}{\rhoAB}$ as well.
\end{lem}

\begin{IEEEproof}
Remember that the generalized fidelity $F(\sigma,\tau)$ is equal to 
$\norm{\sqrt{\sigma}\sqrt{\tau}}_1$ if at least one of the arguments is normalized. Note also 
that every subnormalized operator $\omega'$ can be written as $\omega'=\tr\omega'\cdot\omega$ 
with a normalized operator $\omega$. Let $\omega_\rmAB=\frac{\rhop_\rmAB}{\tr\rhop_\rmAB}$. 
Then 
\begin{equation*}
\begin{split}
F(\rhop_\rmAB,\rhoAB)&=\norm{\sqrt{\rhop_\rmAB}\sqrt{\rhoAB}}_1\\
&=\norm{\sqrt{\tr\rhop_\rmAB\cdot\omega_\rmAB}\sqrt{\rhoAB}}_1\\
&\leq\norm{\sqrt{\omega_\rmAB}\sqrt{\rhoAB}}_1=F(\omega_\rmAB,\rho_\rmAB).
\end{split}
\end{equation*}
Therefore, the purified distance is 
\begin{equation*} 
\begin{split}
P(\rhop_\rmAB,\rhoAB)&=\sqrt{1-F^2(\rhop_\rmAB,\rhoAB)}\\
&\geq\sqrt{1-F^2(\omega_\rmAB,\rhoAB)}\\
&=P(\omega_\rmAB,\rho_\rmAB),
\end{split}
\end{equation*}
which concludes the proof.
\end{IEEEproof}

With this lemma, we can show that there always exists an optimal operator for 
$\Imaxbs{\eps}$ that is normalized.

\begin{lem}\label{30}
Let $\rho\in\Seq{\Hil{AB}}$ and $\eps\geq 0$. Then there exists a normalized 
state $\rhop\in\BAB{\eps}{\rho}$ with $^2I_{\max}(\rmA:\rmB)_{\rhop}=\Imaxbs{\eps}$.
\end{lem}

\begin{IEEEproof}
Let $\bar\rho_\rmAB\in\BAB{\eps}{\rho}$ be any operator satisfying 
$^2I_{\max}(\rmA:\rmB)_{\bar\rho}=\Imaxbs{\eps}$ and let $\smb$ be such that 
\begin{align*}
&2^{\Imaxbs{\eps}}\bar\rho_\rmA\otimes\sigma_\rmB\geq\bar\rho_\rmAB\\
\Rightarrow{} &2^{\Imaxbs{\eps}}\frac{\bar\rho_\rmA}{\tr\bar\rho_\rmAB}\otimes\sigma_\rmB\geq\frac{\bar\rho_\rmAB}{\tr\bar\rho_\rmAB}.
\end{align*}
Hence, 
$$\Imaxbs{\eps}\geq\rendiv{\max}{\frac{\bar\rho_\rmAB}{\tr\bar\rho_\rmAB}}{\frac{\bar\rho_\rmA}{\tr\bar\rho_\rmAB}\otimes\sigma_\rmB},$$ 
but because of Lemma~\ref{10} we find that actually equality holds. We thus 
conclude that if $\bar\rho$ optimizes $\Imaxbs{\eps}$, then so does 
$\rhop=\frac{\bar\rho}{\tr\bar\rho}$.
\end{IEEEproof}

We can prove an analogous and in fact stricter statement about $\Imaxas{\eps}$. 
We give it here for the sake of completeness.

\begin{lem}\label{11}
Let $\rhoAB\in S_=\left(\Hil{AB}\right)$, $\eps\geq 0$ and let 
$\rhop_\rmAB\in\BAB{\eps}{\rhoAB}$ optimize $\Imaxas{\eps}$. Then $\tr\rhop_\rmAB=1.$
\end{lem}

\begin{IEEEproof}
Let $\omega_\rmAB=\frac{\rhop_\rmAB}{\tr\rhop_\rmAB}$. It holds that for 
$k=2^{\Imaxas{\eps}}$
\begin{align*}
&k\cdot\rhop_\rmA\otimes\rhop_\rmB\geq\rhop_\rmAB\\
\Rightarrow {}&k\cdot\tr\rhop_\rmAB\cdot\omega_\rmA\otimes\omega_\rmB\geq\frac{\rhop_\rmAB}{\tr\rhop_\rmAB}=\omega_\rmAB\\
\Rightarrow {}&\Imaxas{\eps}+\log\tr\rhop_\rmAB\geq{^1I_{\max}(\rmA:\rmB)}_\omega,
\end{align*} 
where $\log\tr\rhop_\rmAB\leq 0$. If however this inequality is strict, 
{\it i.e.} $\log\tr\rhop_\rmAB< 0$, this would be a contradiction to the 
optimality of $\rhop_\rmAB$ according to Lemma~\ref{10} and therefore 
$\tr\rhop_\rmAB=1.$
\end{IEEEproof}

\subsection{Proof of Theorem \ref{equiv:1}}

The proof is analogous to the reasoning in Lemma 20 in \cite{TSSR11}. We divide 
it into three steps. Claim \ref{24.1} is a crucial step in the proof of 
Claim \ref{24.2}, from which in turn the result follows.

\begin{claim}\label{24.1}
Let $\rho_\rmABC$ be a purification of $\rhoAB\in S_\leq\left(\Hil{AB}\right)$ and 
$\eps>0$. Then there exists a projector $\Pi_{\rm BC}$ on ${\cal{H}}_{\rm BC}$ such 
that $\tilde{\rho}_\rmABC\defeq\Pi_{\rm BC}\rho_\rmABC\Pi_{\rm BC}\in\BAB{\eps}{\rho_\rmABC}$ 
and 
\begin{equation}
\begin{split}
&\min_{\smb}\rendiv{\max}{\tilde{\rho}_\rmAB}{\rho_\rmA\otimes\sigma_\rmB}\\
\leq{} &\min_{\substack{\sma,\\\smb}}\rendiv{\max}{\rho_\rmAB}{\sigma_\rmA\otimes\sigma_\rmB}+\log{\frac{1}{1-\sqrt{1-\eps^2}}}.
\end{split}
\end{equation} 
\end{claim}

\begin{IEEEproof}
The strategy of the proof is to define $\Pi_{\rm BC}$ as the dual projector 
with respect to $\rho_\rmABC$ (in the sense of Lemma~\ref{prel5}) of a conveniently 
chosen $\Pi_\rmA$ with $\supp\Pi_\rmA\subseteq\supp\rhoA$. Fix 
$\lambda,\bar{\sigma}_\rmA,$ and $\bar{\sigma}_\rmB$ such that 
\begin{equation*}
\begin{split}
&\min_{\substack{\sma,\\\smb}}\rendiv{\max}{\rho_\rmAB}{\sigma_\rmA\otimes\sigma_\rmB}\\
={}&\rendiv{\max}{\rhoAB}{\bar{\sigma}_\rmA\otimes\bar{\sigma}_\rmB}=\log\lambda.
\end{split}
\end{equation*} 
Note that by construction we have $\tilde{\rho}_\rmA\leq\rhoA$ and 
$\supp\tilde{\rho}_\rmB\subseteq\supp\rhoB$, so that we find 
\begin{align*}
\supp\tilde\rho_\rmAB&\subseteq\supp(\tilde\rho_\rmA\otimes\tilde\rho_\rmB)\\
&\subseteq\supp(\rho_\rmA\otimes\rho_\rmB)\\
&\subseteq\supp(\rho_\rmA\otimes\bar\sigma_\rmB).
\end{align*}
Therefore $\rendiv{\max}{\tilde{\rho}_\rmAB}{\rho_\rmA\otimes\bar\sigma_\rmB}$ is 
well defined and we can write 
\begin{equation*}
\begin{split}
&\min_{\smb}\rendiv{\max}{\tilde{\rho}_\rmAB}{\rho_\rmA\otimes\sigma_\rmB}\\
\leq{} &\rendiv{\max}{\tilde{\rho}_\rmAB}{\rho_\rmA\otimes\bar{\sigma}_\rmB}\\
={}&\log\norm{\rhoA^{-\frac{1}{2}}\otimes\bar{\sigma}^{-\frac{1}{2}}_\rmB\tilde{\rho}_\rmAB\rhoA^{-\frac{1}{2}}\otimes\bar{\sigma}_\rmB^{-\frac{1}{2}}}_{\infty}.
\end{split}
\end{equation*} 
Defining $\Pi_\rmA$ as the dual projector of $\Pi_{\rm BC}$ and using the 
inequality $\lambda\bar{\sigma}_\rmA\otimes\bar{\sigma}_\rmB\geq\rhoAB$ we obtain
\begin{align*}
&\norm{\rhoA^{-\frac{1}{2}}\otimes\bar{\sigma}^{-\frac{1}{2}}_\rmB\tilde{\rho}_\rmAB\rhoA^{-\frac{1}{2}}\otimes\bar{\sigma}_\rmB^{-\frac{1}{2}}}_{\infty}\\
={}&\norm{\bar{\sigma}^{-\frac{1}{2}}_\rmB\tr_{\rm C}\left(\rhoA^{-\frac{1}{2}}\otimes\Pi_{\rm BC}{\rho}_\rmABC\rhoA^{-\frac{1}{2}}\otimes\Pi_{\rm BC}\right)\bar{\sigma}_\rmB^{-\frac{1}{2}}}_\infty\\
={}&\norm{\bar{\sigma}^{-\frac{1}{2}}_\rmB\Pi_\rmA\rhoA^{-\frac{1}{2}}{\rho}_\rmAB\rhoA^{-\frac{1}{2}}\Pi_\rmA\bar{\sigma}_\rmB^{-\frac{1}{2}}}_\infty\\
\leq{}&\lambda\norm{\bar{\sigma}^{-\frac{1}{2}}_\rmB\Pi_\rmA\rhoA^{-\frac{1}{2}}\bar{\sigma}_\rmA\otimes\bar{\sigma}_\rmB\rhoA^{-\frac{1}{2}}\Pi_\rmA\bar{\sigma}_\rmB^{-\frac{1}{2}}}_\infty\\
={}&\lambda\norm{\Pi_\rmA\rhoA^{-\frac{1}{2}}\bar{\sigma}_\rmA\rhoA^{-\frac{1}{2}}\Pi_\rmA\otimes\bar{\sigma}_\rmB^0}_\infty\\
={}&\lambda\norm{\Pi_\rmA\Gamma_\rmA\Pi_\rmA}_\infty,
\end{align*}
where we have introduced 
$\Gamma_\rmA\defeq\rhoA^{-\frac{1}{2}}\bar{\sigma}_\rmA\rhoA^{-\frac{1}{2}}$. 
Thus, we find 
\begin{equation*}
\begin{split}
&\min_{\smb}\rendiv{\max}{\tilde{\rho}_\rmAB}{\rho_\rmA\otimes\sigma_\rmB}\\
\leq {}&\min_{\substack{\sma,\\\smb}}\rendiv{\max}{\rho_\rmAB}{\sigma_\rmA\otimes\sigma_\rmB}+\log\norm{\Pi_\rmA\Gamma_\rmA\Pi_\rmA}_\infty.
\end{split}
\end{equation*}
By Lemma~\ref{prel9} it holds that 
\begin{equation}\label{d:54}
\begin{split}
P(\rho_\rmABC,\tilde{\rho}_\rmABC)&\leq\sqrt{2\cdot\tr(\Pi_{\rm BC}^\bot\rho_\rmABC)-\left(\tr(\Pi_{\rm BC}^\bot\rho_\rmABC)\right)^2}\\
&=\sqrt{2\cdot\tr(\Pi_\rmA^\bot\rho_\rmA)-\left(\tr(\Pi_\rmA^\bot\rho_\rmA)\right)^2},
\end{split}
\end{equation} 
where $\Pi^\bot=\id-\Pi$. Now we define $\Pi_\rmA$ to be the minimum rank projector 
on the smallest eigenvalues of $\Gamma_\rmA$ such that 
$\tr(\Pi_\rmA^\bot\rhoA)\leq{1-\sqrt{1-\eps^2}}$. With (\ref{d:54}) this implies 
$P(\rho_\rmABC,\tilde{\rho}_\rmABC)\leq\eps$ since $t\mapsto\sqrt{2t-t^2}$ is 
monotonically increasing in the interval $[0,1]$. It remains to show that with 
our choice of $\Pi_\rmA$ 
$$\norm{\Pi_\rmA\Gamma_\rmA\Pi_\rmA}_\infty\leq\frac{1}{1-\sqrt{1-\eps^2}}$$ holds. 
This, however, can be shown in an identical manner as it is done in the proof of 
Lemma 21 in \cite{TSSR11}. The only difference is that we have chosen $\Pi_\rmA$ 
such that $\tr(\Pi_\rmA^\bot\rhoA)\leq{1-\sqrt{1-\eps^2}}$, instead of 
$\tr(\Pi_\rmA^\bot\rhoA)\leq\frac{\eps^2}{2}$, which eventually leads to slightly 
tighter correction terms.
\end{IEEEproof}

\begin{claim}\label{24.2}
For any $\rhoAB\in S_\leq\left(\Hil{AB}\right)$ there exists a state 
$\bar\rho_\rmAB\in\BAB{\eps}{\rhoAB}$ that satisfies 
\begin{equation}
\begin{split}
&\min_{\smb}\rendiv{\max}{\bar{\rho}_\rmAB}{\bar\rho_\rmA\otimes\sigma_\rmB}\\
\leq{}&\min_{\substack{\sma,\\\smb}}\rendiv{\max}{\rho_\rmAB}{\sigma_\rmA\otimes\sigma_\rmB}+c(\eps,\rhoAB),
\end{split}
\end{equation} 
where $c(\eps,\rhoAB)\defeq\log\left({\frac{1}{1-\sqrt{1-\eps^2}}}+\frac{1}{\tr\rhoAB}\right)$.
\end{claim}

\begin{IEEEproof}
Let $\lambda,\bar{\sigma}_\rmA,\bar{\sigma}_\rmB$ be such that 
\begin{equation*}
\begin{split}
&\min_{\substack{\sma,\\\smb}}\rendiv{\max}{\rho_\rmAB}{\sigma_\rmA\otimes\sigma_\rmB}\\
={}&\rendiv{\max}{\rhoAB}{\bar{\sigma}_\rmA\otimes\bar{\sigma}_\rmB}=\log\lambda.
\end{split}
\end{equation*}
Let us also define the positive semi-definite operator 
$\Delta_\rmA\defeq\rhoA-\tilde\rho_\rmA$ and set 
$\bar\rho_\rmAB=\tilde\rho_\rmAB+\Delta_\rmA\otimes\bar\sigma_\rmB$. It holds that 
$\bar\rho_\rmA=\rhoA$ and $\bar\rho_\rmAB\app{\eps}\rhoAB$, which can be seen 
as follows: since $\tilde\rho_\rmAB\leq\bar\rho_\rmAB$, it also holds that 
$\norm{\sqrt{\tilde\rho_\rmAB}\sqrt{\rhoAB}}_1\leq\norm{\sqrt{\bar\rho_\rmAB}\sqrt{\rhoAB}}_1$. 
Hence,
\begin{equation*}
\begin{split}
F(\bar\rho_\rmAB,\rhoAB)&\geq\norm{\sqrt{\tilde\rho_\rmAB}\sqrt{\rhoAB}}_1+1-\tr\rhoAB\\
&\geq\norm{\sqrt{\tilde\rho_\rmABC}\sqrt{\rho_\rmABC}}_1+1-\tr\rhoAB\\
&=1-\tr(\Pi_{\rm BC}^\bot\rho_{\rm BC})\\
&\geq {\sqrt{1-\eps^2}},
\end{split}
\end{equation*}
and thus $P(\bar\rho_\rmAB,\rhoAB)\leq\eps$. The equality in the penultimate line is a 
consequence of (\ref{fid3}) in Lemma \ref{fid}.\\

Finally, we use $\bar\rho_\rmA=\rhoA$ and 
$\bar\rho_\rmAB\leq\tilde\rho_\rmAB+\rhoA\otimes\bar\sigma_\rmB$ to find
\begin{align*}
&\min_{\smb}\rendiv{\max}{\bar{\rho}_\rmAB}{\bar\rho_\rmA\otimes\sigma_\rmB}\\
\leq{}&\log\norm{\rhoA^{-\frac{1}{2}}\otimes\bar{\sigma}^{-\frac{1}{2}}_\rmB\bar{\rho}_\rmAB\rhoA^{-\frac{1}{2}}\otimes\bar{\sigma}_\rmB^{-\frac{1}{2}}}_{\infty}\\
\leq{}&\log\norm{\rhoA^{-\frac{1}{2}}\otimes\bar{\sigma}^{-\frac{1}{2}}_\rmB(\tilde\rho_\rmAB+\rhoA\otimes\bar\sigma_\rmB)\rhoA^{-\frac{1}{2}}\otimes\bar{\sigma}_\rmB^{-\frac{1}{2}}}_\infty\\
={}&\log\norm{\rhoA^{-\frac{1}{2}}\otimes\bar{\sigma}^{-\frac{1}{2}}_\rmB\tilde\rho_\rmAB\rhoA^{-\frac{1}{2}}\otimes\bar{\sigma}_\rmB^{-\frac{1}{2}}+\rhoA^0\otimes\bar\sigma^0_\rmB}_\infty\\
\leq{}&\log\left(\lambda{\frac{1}{1-\sqrt{1-\eps^2}}}+1\right).
\end{align*}
The first inequality is justified, as 
$$\supp\bar\rho_\rmB=\supp\left(\tilde\rho_\rmB+\tr(\Delta_\rmA)\cdot\bar\sigma_\rmB\right)\subseteq\supp\bar\sigma_\rmB.$$ 
Since $\lambda\geq\tr\rhoAB$, we conclude
\begin{equation*}
\begin{split}
&\min_{\smb}\rendiv{\max}{\bar{\rho}_\rmAB}{\bar\rho_\rmA\otimes\sigma_\rmB}\\
\leq{}&\log\lambda+\log\left({\frac{1}{1-\sqrt{1-\eps^2}}}+\frac{1}{\tr\rhoAB}\right),
\end{split}
\end{equation*}
thus completing the proof of Claim~\ref{24.2}.
\end{IEEEproof}

It is now straightforward to prove the upper bound in the theorem, the lower bound 
given by $$\Imaxcs{\eps+\epsp}\leq\Imaxbs{\eps+\epsp}$$ being clear from the 
definitions. Let $\rhop_\rmAB\in\BAB{\epsp}{\rhoAB}$ be the operator that optimizes 
$\Imaxcs{\epsp}$. Then, by Claim~\ref{24.2}, there exists an operator 
$\bar\rho_\rmAB\in\BAB{\eps+\epsp}{\rhoAB}$ such that 
\begin{align*}
&\min_{\smb}\rendiv{\max}{\bar{\rho}_\rmAB}{\bar\rho_\rmA\otimes\sigma_\rmB}\\
\leq{}&\Imaxcs{\epsp}+\log\left({\frac{1}{1-\sqrt{1-\eps^2}}}+\frac{1}{\tr\rhop_\rmAB}\right)\\
\leq{}&\Imaxcs{\epsp}+\log\left({\frac{1}{1-\sqrt{1-\eps^2}}}+\frac{1}{1-\epsp}\right).
\end{align*}
It remains to notice that by definition of $^2I_{\max}^\eps$ $$\Imaxbs{\eps+\epsp}\leq\min_{\smb}\rendiv{\max}{\bar{\rho}_\rmAB}{\bar\rho_\rmA\otimes\sigma_\rmB},$$ which concludes the proof of Theorem \ref{equiv:1}.

\subsection{Proof of Theorem \ref{equiv:2}}

The derivation of the equivalence between $^2I_{\max}^\eps$ and $^1I_{\max}^\eps$ is 
very similar to the one of Theorem \ref{equiv:1} and is therefore not carried out 
with all of its details here. Again, we only need to prove the upper bound of 
the theorem, since 
$$\Imaxbs{\eps+2\sqrt{\eps}+\epsp}\leq\Imaxas{\eps+2\sqrt{\eps}+\epsp}$$ 
follows directly from the definitions of the quantities. 

We find the following fact, analogous to Claim~\ref{24.1}:

\begin{claim}\label{25.1}
Let $\rho_\rmABC$ be a purification of $\rhoAB\in S_\leq\left(\Hil{AB}\right)$ and $\eps>0$. 
Then there exists a projector $\Pi_{\rm AC}$ on ${\cal{H}}_{\rm AC}$ such that 
$\tilde{\rho}_\rmABC\defeq\Pi_{\rm AC}\rho_\rmABC\Pi_{\rm AC}\in\BAB{\eps}{\rho_\rmABC}$ and 
\begin{equation}
\begin{split}
&\rendiv{\max}{\tilde{\rho}_\rmAB}{\rho_\rmA\otimes\rho_\rmB}\\
\leq {}&\min_{\smb}\rendiv{\max}{\rho_\rmAB}{\rho_\rmA\otimes\sigma_\rmB}+\log{\frac{1}{1-\sqrt{1-\eps^2}}}.
\end{split}
\end{equation}
\end{claim}
\begin{IEEEproof}
The proof of this claim can straightforwardly be adapted from the proof of Claim \ref{24.1}. 
We then end up estimating a term $\norm{\Pi_\rmB\Gamma_\rmB\Pi_\rmB}_\infty$ 
(with $\Gamma_\rmB=\rhoB^{-\frac{1}{2}}\sigma_\rmB\rhoB^{-\frac{1}{2}}$) on system 
$\rmB$ instead of $\rmA$. As before, we can choose $\Pi_{\rm B}$ such that its dual $\Pi_{\rm AC}$ 
satisfies $\tr(\Pi^\bot_{\rm AC}\rho_\rmABC)\leq{1-\sqrt{1-\eps^2}}$.
\end{IEEEproof}

In the following it is sufficient for our purposes to assume that $\rhoAB$ is normalized, 
thanks to Lemma~\ref{30}. Now define $\Delta_\rmABC\defeq\rho_\rmABC-\tilde\rho_\rmABC$ 
and 
$$\bar\rho_\rmAB\defeq k\cdot\left(\tilde\rho_\rmAB+\rhoA\otimes\Delta_\rmB+\Delta_\rmA\otimes\rhoB\right),$$ 
with $k\defeq\frac{1}{1+\tr\Delta_\rmABC}$. Notice that $\Delta_\rmB\geq 0$, but 
$\Delta_\rmA$ and thus $\bar\rho_\rmAB$ is not necessarily positive semi-definite. 
However, $\tr\bar\rho_\rmAB=1$ and by construction we have that $\bar\rho_\rmA=\rhoA$ 
and $\bar\rho_\rmB=\rhoB$. We now want to construct from it a positive semi-definite 
and sub-normalized operator $\hat\rho_\rmAB$ such that $^1I_{\max}(\rmA:\rmB)_{\hat\rho}$ 
is a lower bound to $\rendiv{\max}{\tilde{\rho}_\rmAB}{\rho_\rmA\otimes\rho_\rmB}$ and 
$P(\rhoAB,\hat\rho_\rmAB)\leq c(\eps)$ with $c(\eps)$ a function that vanishes as 
$\eps\rightarrow 0$.

We can write $\Delta_\rmA$ as $\Delta_\rmA=\Delta_\rmA^+-\Delta_\rmA^-$, where 
$\Delta_\rmA^+$ and $\Delta_\rmA^-$ are positive semi-definite operators with 
mutually orthogonal supports. Now we define 
\begin{align*}
\hat\rho_\rmAB&\defeq n\cdot(\bar\rho_\rmAB+k\cdot\Delta_\rmA^-\otimes\rhoB)\\
&=nk\cdot\left(\tilde\rho_\rmAB+\rhoA\otimes\Delta_\rmB+\Delta_\rmA^+\otimes\rhoB\right),
\end{align*}
where $n\defeq(1+k\cdot\tr\Delta_\rmA^-)^{-1}$ is a normalization constant such that 
$\tr\hat\rho_\rmAB=1$. Clearly now, $\hat\rho_\rmAB$ is positive semi-definite and 
we want to estimate $\rendiv{\max}{\hat\rho_\rmAB}{\hat\rho_\rmA\otimes\hat\rho_\rmB}$. 
Notice that $\hat\rho_\rmA=n\cdot(\rhoA+k\Delta_\rmA^-)$ and $\hat\rho_\rmB=\rhoB$. 
Hence,
\begin{align*}
&\rendiv{\max}{\hat\rho_\rmAB}{\hat\rho_\rmA\otimes\hat\rho_\rmB}\\
={}&\log\norm{\hat\rho_\rmA^{-\frac{1}{2}}\otimes\hat\rho^{-\frac{1}{2}}_\rmB\hat\rho_\rmAB\hat\rho_\rmA^{-\frac{1}{2}}\otimes\hat\rho_\rmB^{-\frac{1}{2}}}_\infty\\
={}&\log\Big\Vert(\rhoA+k\da)^{-\frac{1}{2}}\otimes\rhoB^{-\frac{1}{2}}\bar\rho_\rmAB(\rhoA+k\da)^{-\frac{1}{2}}\otimes\rhoB^{-\frac{1}{2}}\\
&~~~~~~~~~~~+(\rhoA+k\da)^{-\frac{1}{2}}k\da(\rhoA+k\da)^{-\frac{1}{2}}\otimes\rho^0_\rmB\Big\Vert_\infty,
\end{align*}
and, with the triangle inequality,
\begin{equation*}
\begin{split}
&\rendiv{\max}{\hat\rho_\rmAB}{\hat\rho_\rmA\otimes\hat\rho_\rmB}\leq\\
&\log\Big(\norm{(\rhoA+k\da)^{-\frac{1}{2}}\otimes\rhoB^{-\frac{1}{2}}\bar\rho_\rmAB(\rhoA+k\da)^{-\frac{1}{2}}\otimes\rhoB^{-\frac{1}{2}}}_\infty\\
&+\norm{(\rhoA+k\da)^{-\frac{1}{2}}k\da(\rhoA+k\da)^{-\frac{1}{2}}}_\infty\Big).
\end{split}
\end{equation*}
We can decompose the first term in the logarithm even further and with $k\leq 1$, $\Delta_\rmB\leq\rhoB$ obtain
\begin{equation*}
\begin{split}
&\rendiv{\max}{\hat\rho_\rmAB}{\hat\rho_\rmA\otimes\hat\rho_\rmB}\leq\\
&\log\Big(\norm{(\rhoA+k\da)^{-\frac{1}{2}}\otimes\rhoB^{-\frac{1}{2}}\tilde\rho_\rmAB(\rhoA+k\da)^{-\frac{1}{2}}\otimes\rhoB^{-\frac{1}{2}}}_\infty\\
&+2\norm{(\rhoA+k\da)^{-\frac{1}{2}}\rhoA(\rhoA+k\da)^{-\frac{1}{2}}}_\infty\\
&+\norm{(\rhoA+k\da)^{-\frac{1}{2}}\tilde\rho_\rmA(\rhoA+k\da)^{-\frac{1}{2}}}_\infty\\
&+\norm{(\rhoA+k\da)^{-\frac{1}{2}}k\da(\rhoA+k\da)^{-\frac{1}{2}}}_\infty\Big).
\end{split}
\end{equation*}
Now we apply Lemma \ref{simple} to all terms inside the logarithm and replace 
$(\rhoA+k\da)$ with $\rhoA$ in the first three terms and with $k\da$ in the last one, 
obtaining
\begin{equation*}
\begin{split}
&\rendiv{\max}{\hat\rho_\rmAB}{\hat\rho_\rmA\otimes\hat\rho_\rmB}\\
\leq{}&\log\left(2\norm{\rhoA^{-\frac{1}{2}}\otimes\rhoB^{-\frac{1}{2}}\tilde\rho_\rmAB\rhoA^{-\frac{1}{2}}\otimes\rhoB^{-\frac{1}{2}}}_\infty+3\right)\\
\leq{}&\rendiv{\max}{\tilde{\rho}_\rmAB}{\rho_\rmA\otimes\rho_\rmB}+\log\left(2+\frac{3}{1-\eps}\right).\\
\end{split}
\end{equation*}
In the last line, we have used that 
$$\norm{\rhoA^{-\frac{1}{2}}\otimes\rhoB^{-\frac{1}{2}}\tilde\rho_\rmAB\rhoA^{-\frac{1}{2}}\otimes\rhoB^{-\frac{1}{2}}}_\infty\geq 1-\eps.$$ 
Thus,
\begin{equation*}
\begin{split}
&\rendiv{\max}{\hat\rho_\rmAB}{\hat\rho_\rmA\otimes\hat\rho_\rmB}\\
\leq{}&\min_{\smb}\rendiv{\max}{\rho_\rmAB}{\rho_\rmA\otimes\sigma_\rmB}\\
&+\log\left({\frac{2(1-\eps)+3}{(1-\eps)(1-\sqrt{1-\eps^2})}}\right).\\
\end{split}
\end{equation*}

Let us finally find an estimate for $P(\rhoAB,\hat\rho_\rmAB)$. Recall that 
$$\tr\Delta_\rmABC=\tr(\Pi^\bot_{\rm AC}\rho_\rmABC)\leq{1-\sqrt{1-\eps^2}}$$ 
according to our choice of $\Pi_{\rm AC}$ in Claim~\ref{25.1}, which implies 
$$k\geq\frac{1}{{2-\sqrt{1-\eps^2}}}.$$ We further have that 
$\tr\Delta_\rmA^-\leq 2\eps$ and therefore $n\geq\frac{1}{1+2\eps}$. 
Thus, with Corollary~\ref{prel11}, 
\begin{equation*}
P(\tilde\rho_\rmAB,nk\cdot\tilde\rho_\rmAB)\leq\sqrt{1-n^2k^2}\leq2\sqrt\eps,
\end{equation*} 
and consequently
\begin{align*}
P(\rhoAB,nk\cdot\tilde\rho_\rmAB)&\leq P(\rhoAB,\tilde\rho_\rmAB)+P(\tilde\rho_\rmAB,nk\cdot\tilde\rho_\rmAB)\\
&\leq \eps+2\sqrt\eps.
\end{align*}
As $nk\cdot\tilde\rho_\rmAB\leq\hat\rho_\rmAB$ and therefore with Lemma \ref{fid} 
$$\norm{\sqrt{nk\cdot\tilde\rho_\rmAB}\sqrt{\rhoAB}}_1\leq\norm{\sqrt{\hat\rho_\rmAB}\sqrt{\rhoAB}}_1,$$ 
we conclude that also $P(\hat\rho_\rmAB,\rhoAB)\leq\eps+2\sqrt{\eps}$. 

In summary, we have just proven the following claim.
\begin{claim}\label{25.2}
For any $\rhoAB\in S_=\left(\Hil{AB}\right)$ and $\eps>0$, there exists a state 
$\hat\rho_\rmAB\app{\eps+2\sqrt{\eps}}\rhoAB$ that satisfies 
\begin{equation}
\begin{split}
\rendiv{\max}{\hat{\rho}_\rmAB}{\hat\rho_\rmA\otimes\hat\rho_\rmB}\leq&\min_{\smb}\rendiv{\max}{\rho_\rmAB}{\rho_\rmA\otimes\sigma_\rmB}\\
&+\log\left({\frac{2(1-\eps)+3}{(1-\eps)(1-\sqrt{1-\eps^2})}}\right).
\end{split}
\end{equation}
\end{claim}

To conclude the proof of Theorem \ref{equiv:2}, let $\rhoAB\in\Seq{\Hil{AB}}$ and 
let $\rhop_\rmAB\in\BAB{\epsp}{\rhoAB}$ be a normalized operator such that 
$\Imaxbs{\epsp}={^2I}_{\max}(\rmA:\rmB)_{\rhop}$. Applying Claim~\ref{25.2} to 
$\rhop_\rmAB$ yields the result.

\end{document}